\documentclass[aps,pra,preprint,floatfix]{revtex4}
\usepackage{graphicx}
\usepackage{multirow}

\begin{document}

\title{Stimulated Raman scattering in an optical parametric oscillator based on periodically poled MgO-doped stoichiometric LiTaO$_3$}

\author{T-H. My, O. Robin, O. Mhibik, C. Drag, and F. Bretenaker}

\address{Laboratoire Aim\'e Cotton , CNRS-Universit\'e Paris Sud 11,\\ Campus d'Orsay, 91405 Orsay Cedex, France}

\date{20 February 2009}



\begin{abstract}  The evolution versus pump power of the spectrum of a singly resonant optical parametric oscillator based on an MgO-doped periodically
poled stoichiometric lithium tantalate crystal is observed. The onset of cascade Raman lasing due to stimulated Raman scattering in the nonlinear crystal is
analyzed. Spurious frequency doubling and sum-frequency generation phenomena are observed and understood. A strong reduction of the intracavity Raman
scattering is obtained by a careful adjustment of the cavity losses.
\end{abstract}

\maketitle

\section{Introduction}
Continuous wave optical parametric oscillators (OPOs) are important sources of high power, widely tunable, narrowband radiation, for applications in
spectroscopy and sensing. Such sources, when based on periodically poled MgO-doped stoichiometric LiTaO$_3$ (MgO-PPSLT), are very promising because of the
increased resistance of this crystal to the photorefractive effect, its large optical nonlinearity, and its high optical damage threshold \cite{Strossner1999,
Melkonian2007, Samanta2007, Samanta2008-1}. In cw Singly Resonant OPOs (SROPOs) developed to date, oscillation of a single frequency has been observed at low
pumping levels (up to 5 times threshold) due to the relatively high threshold of SROPOs (a few Watts level). Exploring theoretically much larger values of the
relative pumping ratio in cw SROPOs, Kreuzer has predicted that single frequency oscillation should occur at moderate pumping ratios, while multimode
oscillation must occur above a critical value of the pumping ratio \cite{Harris1969}. This has been experimentally demonstrated in the case of SROPOs based on
periodically poled LiNbO$_3$ \cite{Okishev2006, Henderson2007,Vainio2009} in which the ratio of pump power to oscillation threshold could reach to 16 times.
Moreover, in the same papers \cite{Okishev2006, Henderson2007,Vainio2009}, it has been shown that further increasing the pump power can lead to the outbreak of
new frequencies due to Raman scattering in the PPLN crystal. But till now, no demonstration of such effects has been performed in the case of PPSLT.

\indent Moreover, in recents years, in order to reach wavelengths inaccessible to commercial solid-state lasers, cw intracavity-frequency-doubled SROPOs based
on PPSLT crystals have been developed \cite{Samanta2008-2, My2008}. To be efficient, such systems require a high intracavity signal or idler power. However,
such a requirement may change the spectral properties of these systems. So, for further scaling of their power while retaining a good spectral purity, it is
important to investigate the effects induced by strong intracavity powers in such systems. In particular in this paper, we report on simultaneous parametric
oscillation and stimulated Raman scattering in a MgO-PPSLT based SROPO. We  also describe the spectral characteristics observed at high (up to 15 times
threshold) pumping ratios.
\begin{figure}[htb]
\centering\includegraphics[width=8cm]{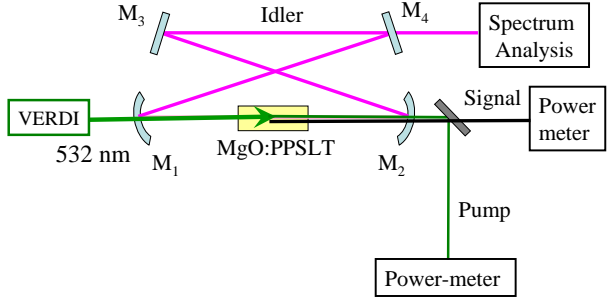} \caption{\label{Fig1} Schematic of SROPO configuration.}
\end{figure}\section{Experimental setup}
Our experimental setup is shown in Fig.\,\ref{Fig1}. As a pump source, we use a cw Verdi laser that produces 10 W of single-frequency radiation at 532 nm. The
nonlinear crystal is a 30-mm-long 1\% MgO-doped PPSLT crystal ($d_{\mathrm{eff}}\simeq11\ \mathrm{pm/V}$) manufactured and coated by HC Photonics. This crystal
contains a single grating with a period of 7.97 $\mu$m. It is designed to lead to quasi-phase-matching conditions for an idler wavelength in the 1200-1400 nm
range, as already shown in Refs. \cite{Samanta2007, Samanta2008-1}. It is antireflection coated for the pump, signal, and idler wavelengths. It is mounted in a
temperature controlled oven, allowing coarse wavelength tuning. The crystal temperature stability is 0.01$^\circ$C. It can be varied between 30 and
230$^\circ$C.

\indent The OPO cavity is a ring cavity and consists in four mirrors. Mirrors M$_1$ and M$_2$ have a 150 mm radius of curvature, and mirrors M$_3$ and M$_4$
are plane. It is resonant for the idler wavelength only. The estimated waist of the idler beam at the middle of the PPSLT crystal is 37 $\mu$m. The pump beam
is focused to a 53 $\mu$m waist inside the PPSLT crystal. All mirrors are designed to exhibit a reflectivity larger than 99.8\% between 1.2 $\mu$m and 1.4
$\mu$m and a transmission larger than 95\% at 532 nm and between 850 nm and 950 nm. The signal and pump output beams are separated by a dichroic mirror as
shown in Fig.\,\ref{Fig1}. Pump depletion is measured by observing the pump power transmitted through the OPO, with and without blocking one leg of the optical
cavity.
\section{Raman laser: results et discussions}
\begin{figure}[htb]
\centering\includegraphics[width=7cm]{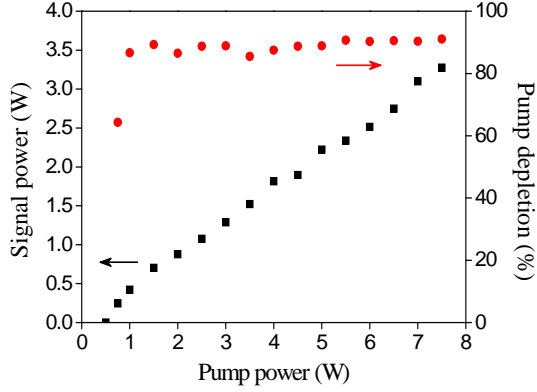} \caption{\label{Fig2} Measured evolutions of the pump depletion and signal power as a function of the pump
power. The cw SROPO  cavity consists of four highly reflecting mirrors at the idler wavelength.}
\end{figure}We obtain the oscillation of the SROPO with a threshold pump power of 500 mW at a crystal temperature of 103$^\circ$C (see Fig.\,\ref{Fig2}). This threshold
value is at least five times lower than the one previously demonstrated in similar cw OPO systems based on PPSLT crystal \cite{Samanta2007}. This is due to the
low round trip losses in our optical cavity, and also to the rather good mode matching of the pump and idler beams. With 7.6 W of available pump power, we are
thus able to pump the OPO at 15 times its oscillation threshold. The evolution of the non resonating signal power and of the pump depletion versus pump power
are reproduced in Fig.\,\ref{Fig2}. By varying the temperature of crystal, the wavelengths can be tuned between 1170 and 1355 nm for the idler, and 876 and
975~nm for the signal. We measure a pump depletion as high as 90\%, as shown on Fig.\,\ref{Fig2}. It can be seen that we obtain 3 W of signal at a maximum pump
power of 7.6 W.

\begin{figure}[htb]
\centering\includegraphics[width=13cm]{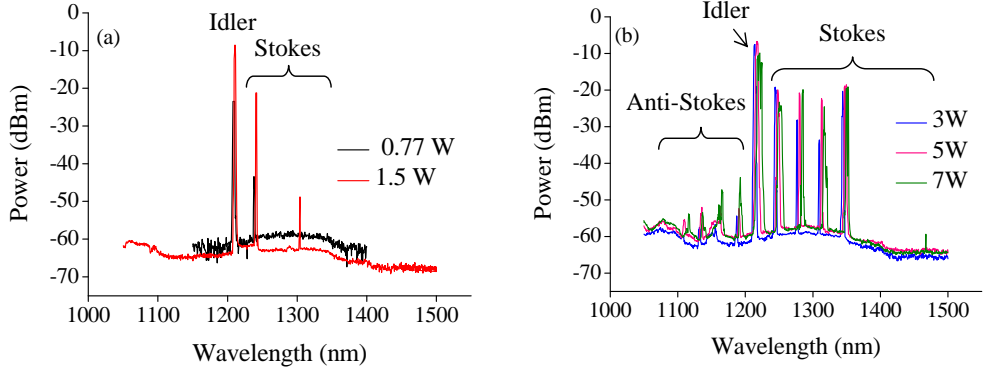} \caption{\label{Fig3} Spectrum of the combined OPO and Raman oscillation at differents pump powers.
Resolution: 2.0 nm.}
\end{figure}
\begin{figure}[htb]
\centering\includegraphics[width=13 cm]{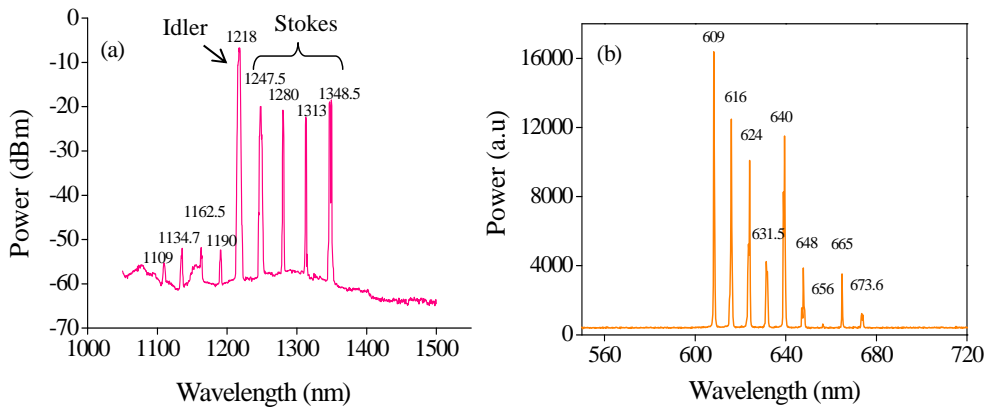} \caption{\label{Fig4} OPO output spectrum for a 5~W pump power, corresponding to 10 times threshold. (a)
Infrared part of the spectrum. Resolution: 1.0 nm. (b) Visible part of the spectrum. Resolution: 0.3 nm.}
\end{figure}

In the following, all the experimental results on the OPO idler spectrum have been obtained at an oven temperature of 103$^\circ$C. The idler spectrum is
measured using an optical spectrum analyzer (model AQ6317B manufactured by ANDO). Fig.\,\ref{Fig3} shows the recorded idler spectrum for five different values
of the pump power. We observe that at low pump powers (close to the OPO threshold), the idler spectrum consists in a single peak. When we increase the pump
power, multiple Raman lines appear in the idler spectrum, evidencing Raman gain lasing in the OPO cavity. The threshold of this Raman laser effect corresponds
to a pump power of 770 mW [see the black spectrum in Fig.\,\ref{Fig3}(a)]. From the measurement of the idler output power and the measured transmission of
mirror M$_4$ ($5\times 10^{-5}$), the intracavity idler power corresponding to the threshold of the Raman laser effect is estimated to be equal to 30 W. The
number and intensity of the Raman lines increases with the pump power, showing a cascade of Raman laser effects. For a pump power up to 3 W, only Stokes lines
appear in the spectrum. For larger pump powers, the spectrum also contains some very weak anti-Stokes lines [see Fig.\,\ref{Fig3}(b)]. To measure the Raman
shift, we can use for example the idler spectrum corresponding to a pump power of 5 W, i. e., a relative pumping level of 10 [see Fig.\,\ref{Fig4}(a)]. From
this spectrum, we find that all the lines are consistent with a Raman shift of around 200 cm$^{-1}$ which corresponds to a phonon mode of the LiTaO$_3$ crystal
\cite{Johnston1968, Penna1976,Boyd2008}.

The threshold of the Raman laser effect corresponds to an intracavity idler power of 30~W, which is equivalent to an intensity of the order of 0.7~MW/cm$^2$ in
the PPSLT crystal. If we suppose that the Raman gain in the crystal is equivalent to the one found in the literature \cite{Boyd2008} for LiTaO$_3$, i. e.,
4.4~cm/GW, this corresponds to a gain of about 0.9\% per round-trip at the Raman shifted wavelength. This is consistent with the order of magnitude of the
round-trip cavity losses between 1200 and 1400~nm.

\begin{table}[h]
\centering\caption{\label{Table1} Explanation of the origin of the different peaks at wavelengths $\lambda_3$ emitted in the visible [see Fig.\,\ref{Fig4}(b)].
The infrared wavelengths $\lambda_1$ and $\lambda_2$ are taken from Fig.\,\ref{Fig4}(a). SHG holds for Second Harmonic Generation and SFG for Sum Frequency
Generation.}
\begin{tabular}{|c|c|c|c|c|}
\hline
$\lambda_3$ (nm) & Mechanism & Relation & $\lambda_1$ (nm) & $\lambda_2$ (nm) \\
\hline
609 & SHG & $\displaystyle{1/\lambda_3=2/\lambda_1}$ & 1218 & $--$ \\
\hline
616 & SFG & $\displaystyle{1/\lambda_3=1/\lambda_1+1/\lambda_2}$ & 1218 & 1247.5 \\
\hline
\multirow{2}{*}{624} & SHG & $\displaystyle{1/\lambda_3=2/\lambda_1}$ & 1247.5 & $--$ \\
  & SFG & $\displaystyle{1/\lambda_3=1/\lambda_1+1/\lambda_2}$ & 1218 & 1280 \\
  \hline
\multirow{2}{*}{631.5} & SFG & $\displaystyle{1/\lambda_3=1/\lambda_1+1/\lambda_2}$ & 1247.5 & 1280 \\
  & SFG & $\displaystyle{1/\lambda_3=1/\lambda_1+1/\lambda_2}$ & 1218 & 1313 \\
  \hline
\multirow{3}{*}{640} & SHG & $\displaystyle{1/\lambda_3=2/\lambda_1}$ & 1280 & $--$ \\
  & SFG & $\displaystyle{1/\lambda_3=1/\lambda_1+1/\lambda_2}$ & 1247.5 & 1313 \\
  & SFG & $\displaystyle{1/\lambda_3=1/\lambda_1+1/\lambda_2}$ & 1218 & 1348.5 \\
\hline
\multirow{2}{*}{648} & SFG & $\displaystyle{1/\lambda_3=1/\lambda_1+1/\lambda_2}$ & 1280 & 1313 \\
  & SFG & $\displaystyle{1/\lambda_3=1/\lambda_1+1/\lambda_2}$ & 1247,5 & 1348,5 \\
\hline
\multirow{2}{*}{656} & SHG & $\displaystyle{1/\lambda_3=2/\lambda_1}$ & 1313 & $--$ \\
   & SFG & $\displaystyle{1/\lambda_3=1/\lambda_1+1/\lambda_2}$ & 1280 & 1348,5 \\
\hline
665 & SFG & $\displaystyle{1/\lambda_3=1/\lambda_1+1/\lambda_2}$ & 1313 & 1348.5 \\
\hline
673.6 & SHG & $\displaystyle{1/\lambda_3=2/\lambda_1}$ & 1348.5 & $--$ \\
\hline
\end{tabular}
\end{table}
Investigation of the OPO output spectrum also revealed unusual peaks in the visible. Fig.\,\ref{Fig4} shows the infrared and visible parts of the output
spectrum for a pump power of 5 W. The visible spectrum was measured using an AvaSpec-2048-2 spectrometer. The origin of these visible wavelengths is summarized
in Table \ref{Table1}. They all come from the second order nonlinearity of the PPSLT crystal, which mixes the infrared frequencies, either by sum frequency
generation or by second harmonic generation. In such a situation, rather than emitting just the idler wavelength, one can see that the OPO emits the idler
frequency, four Stokes Raman shifted wavelengths, four anti-Stokes Raman shifted wavelengths, and nine different wavelengths in the visible. This is too
complicated for applications.

\section{Single-frequency operation}
\begin{figure}[htb]
\centering\includegraphics[width=13cm]{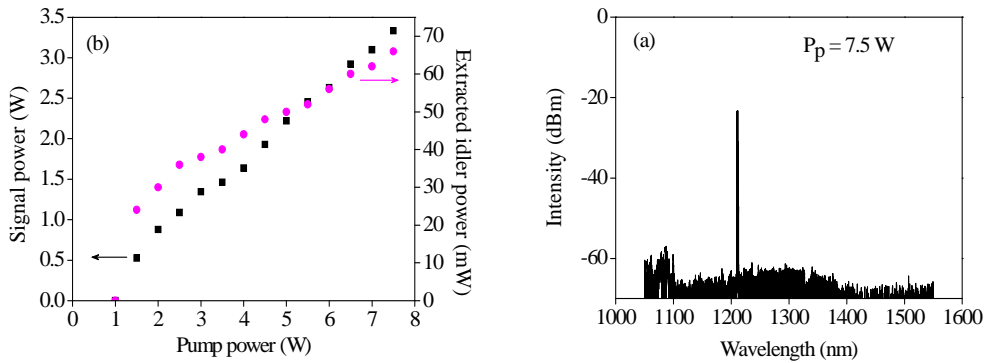} \caption{\label{Fig5} For the configuration where the cavity contains one output coupling mirror (a) Signal
power and extracted idler power as a function of pump power (b) Idler spectrum at maximum pump power of 7.5 W. Resolution: 0.1 nm}
\end{figure}

\indent Indeed, for the applications mentioned in the introduction, we need a source operating in single frequency regime. We thus have to avoid the Raman
laser effect. We then reduce the intracavity idler power and also increased the cavity losses at Raman wavelengths by changing mirror M$_4$ for a mirror with a
transmission of the order of $8\times 10^{-4}$ around 1200~nm. With this mirror, the SROPO threshold is increased to 1.0 W, showing that we have roughly
multiplied the losses for one round-trip in the cavity at the idler wavelength by a factor of 2. With this mirror, the evolutions of the signal and idler
output powers are displayed in Fig.\,\ref{Fig5}(a). The Raman laser threshold is now measured to correspond to a pump power of 5~W, which corresponds to an
idler output power of about 50~mW. This is equivalent to an intracavity idler power of about 60~W. This shows that, in terms of idler power the Raman laser
threshold is now twice larger than with the preceding cavity. This is consistent with the fact that the cavity losses are twice larger than before and that the
Raman gain is proportional to the pump laser intensity \cite{Boyd2008}.

The good thing for applications is that even at higher pump powers, the Raman effect remains very weak and can hardly be detected [see for example the spectrum
of Fig.\,\ref{Fig5}(b), obtained for a pump power of 7.5~W], even though the system is again very efficient, with a pump depletion measured to be equal to
90\%.
\begin{figure}[htb]
\centering\includegraphics[width=13cm]{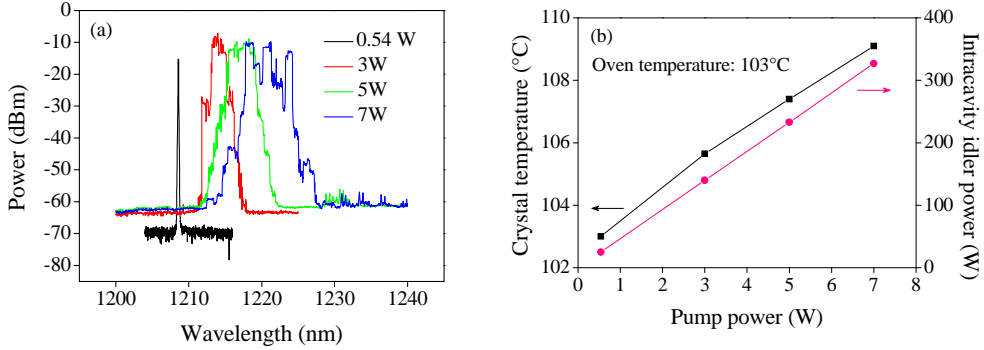} \caption{\label{Fig6} (a) Idler spectrum as a function of pump power. Resolution: 1.0 nm (b) Evolution of the
crystal temperature, deduced from the idler wavelength, and of the intracavity idler power versus pump power. These results were obtained with the four highly
reflecting mirrors.}
\end{figure}

If now we focus on the position and width of the idler frequency peak, some effects occurring at high pump powers can also be observed. Indeed, in the case of
the cavity built with four highly reflecting mirrors, we also observe significant shifts in idler wavelength as a function of the pump power. This effect,
which arises from the heating of the PPSLT crystal, has been previously observed in PPLN \cite{Henderson2006,Vainio2009} and also more recently in PPSLT
\cite{Samanta2008-2}. Fig.\,\ref{Fig6}(a) reproduces a zoom on the idler frequency spectra at different pump powers. At a pump power of 7W (respectively
540~mW), we measure an idler wavelength of 1221 nm (respectively 1208.5~nm). Using the Sellmeier equation for stoichiometric LiTaO$_3$ \cite{Bruner2003}, we
can extract the crystal temperature these wavelengthes. Fig.\,\ref{Fig6}(b) displays the result of this calculation and also the intracavity idler power versus
pump power. It indicates a rise of 6$^\circ$C of the crystal temperature with increased pump power. Since the green-induced infrared absorption is expected to
be low in MgO-PPSLT \cite{Hirohashi2007}, this effect may be due to the absorption of the intracavity idler, and also absorption of the pump and signal in the
MgO-PPSLT crystal. Morever, as the pump power increases, Fig.\,\ref{Fig6}(a) shows that the idler spectrum broadens. This effect has also been observed in the
case of the PPLN crystal \cite{Henderson2007}. Here, we find that, unlike in PPLN, the spectrum showed a symmetric pattern of side modes at all pumping ratios,
up to 15 times threshold.

\section{Conclusion}
In conclusion, we have observed the onset of lasing due stimulated Raman scattering in a cw singly resonant optical parametric oscillator pumped in the green
and based on MgO-PPSLT, with a sub-Watt pump power threshold. This effect has been shown to be due to the high intracavity idler power and the high finesse of
cavity at the Raman shifted wavelengths. We have shown that single wavelength operation of the SROPO can be obtained while keeping several Watts of signal
power and 90\% pump depletion, by optimizing the cavity losses for the idler wavelength. Effects of the crystal heating due to the high intracavity power have
also been observed.

\section*{Acknowledgments}
 This work was partially supported by the Triangle de la Physique. The authors are happy to thank M. Lefebvre, A. Godard, M. Raybaut, and B. Hardy for helpful
 discussions.


\begin{thebibliography}{99}

\bibitem{Strossner1999} U. Str\"ossner, A. Peters, J. Mlynek, and S. Schiller, J.-P. Meyn, and R. Wallenstein,
``Single-frequency continuous-wave radiation from 0.77 to 1.73 $\mu$m generated by a green-pumped optical parametric oscillator with periodically poled
LiTaO$_3$," Opt. Lett. {\bf 24}, 1602-1604 (1999).

\bibitem{Melkonian2007} J.-M. Melkonian, T.-H. My, F. Bretenaker, and C. Drag, ``High spectral purity and tunable operation of a continuous singly
 resonant optical parametric oscillator emitting in the red," Opt. Lett. {\bf 32}, 518-520 (2007).

\bibitem{Samanta2007} G.K. Samanta, G. R. Fayaz, and M. Ebrahim-Zadeh, ``1.59 W, single-frequency, continuous-wave optical parametric oscillator
based on MgO:sPPLT," Opt. Lett. {\bf 32}, 2623-2625 (2007).

\bibitem{Samanta2008-1} G.K. Samanta, and M. Ebrahim-Zadeh, ``Continuous-wave singly-resonant optical parametric oscillator with resonant wave
coupling," Opt. Expr. {\bf 16}, 6883-6888 (2008).

\bibitem{Harris1969} S. E. Harris, ``Tunable optical parametric oscillators," Proc. IEEE {\bf 57}, 2096-2113 (1969).

\bibitem{Okishev2006} A. V. Okishev and J. D. Zuegel, ``Intracavity-pumped Raman laser action in a mid-IR, continuous-wave (cw)
MgO:PPLN optical parametric oscillator,'' Opt. Expr. {\bf 14}, 12169-12173 (2006).

\bibitem{Henderson2007} A. Henderson and R. Stafford, ``Spectral broadening and stimulated Raman conversion in a continuous-wave optical
parametric oscillator," Opt. Lett. {\bf 32}, 1281-1283 (2007).

\bibitem{Vainio2009} M. Vainio, J. Peltola, S. Persijn, F. J. M. Harren, and L. Halonen, ``Thermal effects in singly resonant continuous-wave optical
parametric oscillators,'' Appl. Phys. B {\bf 94}, 411-427 (2009).

\bibitem{Samanta2008-2} G. K. Samanta and M. Ebrahim-Zadeh, ``Continuous-wave, single-frequency, solid-state blue source for the
425-489 nm spectral range," Opt. Lett. {\bf 33}, 1228-1230 (2008).

\bibitem{My2008} T.-H. My, C. Drag, and F. Bretenaker, ``Single-frequency and tunable operation of a continuous
intracavity-frequency-doubled singly resonant optical parametric oscillator," Opt. Lett. {\bf 33}, 1455-1457 (2008).

\bibitem{Johnston1968} W. D. Johnston, Jr. and I. P. Kaminow, ``Temperature dependence of Raman and Rayleigh scattering
 in LiNbO$_3$ and LiTaO$_3$,'' Phys. Rev. {\bf 168,} 1045-1054 (1968).

\bibitem{Penna1976} A. F. Penna, A. Chaves,  P. da R. Andrade, and S. P. S. Porto, ``Light scattering by lithium
tantalate at room temperature,'' Phys. Rev. B {\bf 13,} 4907-4917 (1976).

\bibitem{Boyd2008} R. W. Boyd, \textit{Nonlinear Optics}, 3rd edition (Academic Press, 2008).

\bibitem{Henderson2006} A. Henderson and R. Stafford, ``Intra-cavity power effects in singly resonant cw OPOs,'' Appl. Phys. B. {\bf 85,} 181-184 (2006).

\bibitem{Bruner2003} A. Bruner, D. Eger, M. B. Oron, P. Blau, M. Katz, and S. Ruschin, ``Temperature-dependent Sellmeier equation
 for the refractive index of stoichiometric lithium tantalate," Opt. Lett. {\bf 28}, 194-196 (2003).

\bibitem{Hirohashi2007} J. Hirohashi, V. Pasiskevicius, and F. Laurell, ``Picosecond blue-light-induced
infrared absorption in single-domain and periodically polled ferroelectrics," J. Appl. Phys. {\bf 101}, 033105 (2007).

\end{thebibliography}
\end{document}